# Parameterized Post-Newtonian Orbital Effects in Extrasolar Planets

Lin-Sen Li
Department of Physics, Northeast Normal University,Changchun, 130024, China [ e-mail:dbsd_lls@yahoo.com.cn ]

Abstract: Perturbative Post-Newtonian variations of the standard osculating orbital elements are obtained by using the two-body equations of motion in the Parameterized Post-Newtonian theoretical framework. The results obtained are applied to the Einstein and. Brans - Dicke theories. As a results, the semi-major axis and eccentricity exhibit periodic variation, but no secular changes.. The longitude of periastron and mean longitude at epoch experience both secular and periodic shifts. The Post-Newtonian effects are calculated and discussed for six extrasolar planets .

## I. Introduction

At present, the Post-Newtonian effect has been exhibited gradually in the wake of unceasing development in the Post Newtonian celestial mechanics and due to that the accurate degree of astronomical instruments is heightened unceasingly.. Hence some authors devoted to the research on the subject and scopes, such as Estabrook (1969), Nordtvedt ( 1970 ), Rubincam ( 1977 ), Brumberg (1982,1985,2010), Damour (1985) , Soffel et al (1987), Soffel (1989 ), Klioner & Kopejkin (1992), Calura et al ( 1997), Brumberg et al

(1995 ), Brumberg et al (2001), Iorio ( 2005a, 2005b, 2007a，2011a ), Will ( 2008 ) Everitt et al ( 2011), Kopeikin et al (2011) and Iorio et al, ( 2011). In the Post-Newtomian celestial mechanics there are some best methods One of the best methods is the method of parameterized post-Newtonian Formalism ( PPN method ). because the theories include the various different gravitational theories with different parameters, such as Einstein, Brans-Dicke and other theories.. Hence some authors devoted to the research on this scope, such as, Misner et al, ( 1973 ), , Nordtvedt (1976), Sarmiento (1982), Will (1981, 2006) Moreover, the Asymptoti method ( Brumberg & Kopejkin ,1989, 1990 ) and D S X method ( Damour et al, 1991,1992 ) are also desirable

At present, some authors not only studied the Post-Newtonian effect on the motion of celestial objects in the solar system, but also in the extrasolar planetary system. It is interesting and significant for studying the Post-Newtonian effects on the extrasolar planets because in the extrasolar planetary system the separation between planets and primary star is nearer mutually and planet mass is nearly Jupiter mass. Hence the Post-Newtonian effect on the orbital elements of extraplanets is larger. In the recent years some authors studied the Post-Newtonian effect or the relativistic effect in the extrasolar planets ( Calura & Montanari (1999), Miralda-Escudė. (2002), Wittenmyer et al (2005), Iorio ( 2006, 2011b, 2011c ), Adams & Laughlin ( 2006a, 2006b, 2006c ), Heyl & Gladman (2007), Pál & Kocsis (2008) and Jordăn & Bacos (2008), Ragozzine & Wolf (2009) However, these authors used the method of the general relativity or the post-Newtonian approximation to study this problem. This paper used the parameterized Post-Newtonian theories to study and calculate the Parameteried Post-Newtonian effect on the Extrasolar planets with large eccentric orbit.

## 2. R, S and W components for the parameterized Post-Newtonian perturbing acceleration in the two-body problem

The relative acceleration of two-body with the Post-Newtonian Parameters is given by Will (1981)

$$\vec{a}_{PN} = \frac{m\vec{X}}{r^3}[(2\gamma + 2\beta)\frac{m}{r} - \gamma v^2 + (2 + \alpha_1 - 2\zeta_2)\frac{\mu}{r} -$$

$$-\frac{1}{2}(6 + \alpha_1 + \alpha_2 + \alpha_3)\frac{\mu}{m}v^2 + \frac{3}{2}((1 + \alpha_2)\frac{\mu}{m}(\vec{v}.\vec{n})^2]$$

$$+\frac{m(\vec{X}.\vec{v})\vec{v}}{r^3}[(2\gamma + 2) - \frac{\mu}{m}(2 - \alpha_1 + \alpha_2)]. \qquad (1)$$

Here $m = m_1 + m_2$ $\mu = \frac{m_1 m_2}{m}$, $\widetilde{N} = \frac{\vec{X}}{r}$,

$\vec{X} = \vec{N}.r = \vec{r}$, $\vec{r}.\dot{\vec{t}} = r.\dot{r}$ , $\vec{v} = \dot{\vec{r}} = \dot{r}\vec{N} + r\dot{f}\vec{\lambda}$, $v^2 = \dot{r}^2 + r^2\dot{f}^2$

$$(\vec{v}.\vec{N})^2 = (\vec{v}.\frac{\vec{X}}{r})^2 = \frac{(\vec{v}.\vec{N}r)^2}{r^2} = \frac{(\dot{r}^2.\vec{r})^2}{r^2} = \dot{r}^2$$

$$\frac{m(\vec{X}.\vec{v})\vec{v}}{r^3} = \frac{m}{r^3}(\vec{N}r.\vec{v})\vec{v} = \frac{m}{r^3}(\vec{r}.\dot{\vec{r}})\dot{\vec{r}} = \frac{m}{r^3}(r.\dot{r})(\dot{r}\vec{N} + r\dot{f}\vec{\lambda}) = r\dot{r}\vec{N} + r^2\dot{r}\dot{f}\vec{\lambda}.$$

Here $f$ denotes the true anomaly. $\vec{N}$ is a unit vector in the radial direction and $\vec{\lambda}$ are unit vectors in the orbital plane. $\vec{N}$ is directed along the radial direction, while $\vec{\lambda}$ is perpendicular to $\vec{N}$. In the equation m denotes *G*m and the right side should multipled by $c^{-2}$. *G* is the gravitional constant and *c* is the speed of light. The equation (1) can be written as

$$\vec{a}_{ppn} = \frac{m}{r^3}(r\vec{N})\langle(2\gamma + 2\beta)\frac{m}{r} + (2 + \alpha_1 - 2\varsigma_2)\frac{\mu}{r} - \{(\gamma + \frac{1}{2}(6 + \alpha_1 + \alpha_2 + \alpha_3)\frac{\mu}{m}]\times$$

$$(\dot{r}^2 + r^2\dot{f}^2) + \frac{3}{2}(1 + \alpha_2)\frac{\mu}{m}\dot{r}^2\rangle +$$

$$+\frac{m}{r^3}[(r\dot{r}^2\vec{N} + r^2\dot{r}\dot{f}\vec{\lambda})\{(2\gamma + 2) - (2 - \alpha_1 + \alpha_2)\frac{\mu}{m}\}] \qquad (2)$$

Here $\rightarrow$ denotes vector

We resolve the acceleration $\vec{a}$ into a radial component R$\vec{N}$, a component

S $\vec{\lambda}$, normal to R $\vec{N}$ and a component W normal to the orbital plane.

i.e, $\vec{a} = R\vec{N} + S\vec{\lambda} + W(\vec{N} \times \vec{\lambda})$ $\quad \vec{N} \times \vec{\lambda} = \vec{L}$ ( the unit vector normal to the orbital plane ).

On comparison with the expression (2), we get three scalar accelerative components R, S and W

$$R = \frac{m}{r^2}\left\langle \begin{matrix} (2\gamma + 2\beta)\frac{m}{r} + (2 + \alpha_1 - 2\zeta_2)\frac{\mu}{r} - [\gamma + \frac{1}{2}(6 + \alpha_1 + \alpha_2 + \alpha_3)\frac{\mu}{m}](\dot{r}^2 + r^2 \dot{f}^2) + \\ \frac{3}{2}(1 + \alpha_2)\dot{r}^2 \frac{\mu}{m} + \end{matrix} \right\rangle$$

$$+ + \frac{m}{r^2}\dot{r}^2[\{2\gamma + 2) - (2 - \alpha_1 \ + \alpha_2)\frac{\mu}{m}]$$

$$S = \frac{m}{r}[(2\gamma + 2) - (2 - \alpha_1 + \alpha_2)\frac{\mu}{m}]\dot{r}\dot{f} \tag{3}$$

W = 0

Substituting the following formulas of the problem of two body into the above formula ( Smart, 1953)

$$\dot{f} = \frac{df}{dt} = na^2\sqrt{1 - e^2} / r^2, \quad \dot{r} = \frac{nae \sin f}{\sqrt{1 - e^2}}. \tag{4}$$

$$n^2 a^3 = m.$$

We obtain

$$R = \frac{m^2}{r^2}\left( \frac{2(\gamma + \beta) + (2 + \alpha_1 - 2\zeta_2)\frac{\mu}{m}}{r} + [(\gamma + 2) - \frac{1}{2}(7 - \alpha_1 + \alpha_3)\frac{\mu}{m}]\frac{e^2 \sin^2 f}{p} - \right.$$

$$\left. - [\gamma + \frac{1}{2}(6 + \alpha_1 + \alpha_2 + \alpha_3)\frac{\mu}{m}]\frac{p}{r^2} \right]$$

$$S = \frac{m^2}{r^3}[(2\gamma+2) - (2-\alpha_1+\alpha_2)\frac{\mu}{m}]e\sin f \quad , \qquad (5)$$
$$W = o$$

where $p = a(1-e^2)$.

We can write simply the above formulas as the following formulas:

$$\begin{cases} r^2 R = (\frac{K_1}{r} + K_2\frac{e^2}{p}\sin^2 f + K_3\frac{p}{r^2}), \\ r^2 S = \frac{K_4}{r} e\sin f, \\ W = 0. \end{cases} \qquad (6)$$

Substituting $r = p/(1+e\cos f)$ into the right hand sides of the expression (2), we obtain

$$\begin{cases} r^2 R = \frac{1}{p}\left\{[K_1 + \frac{1}{2}e^2 K_2 + (1+\frac{1}{2}e^2)K_3] + (K_1 + 2K_3)e\cos f + \frac{1}{2}(K_3 - K_2)e^2\cos 2f\right\}, \\ r^2 S = \frac{1}{p}[K_4(1+e\cos f)e\sin f = \frac{c^2}{p}(K_4 e\sin f + \frac{1}{2}K_4 e^2\sin 2f), \\ W = 0. \end{cases} \qquad (7)$$

In it, it is

$$\begin{cases} K_1 = (2\gamma + 2\beta)m^2 + (2 + \alpha_1 - 2\zeta_2)m_1 m_2, \\ K_2 = (\gamma+2)m^2 - \frac{1}{2}(7 - \alpha_1 + \alpha_3)m_1 m_2, \\ K_3 = -\gamma m^2 - \frac{1}{2}(6 + \alpha_1 + \alpha_2 + \alpha_3)m_1 m_2, \\ K_4 = (2\gamma+2)m^2 - (2 - \alpha_1 + \alpha_3)m_1 m_2. \end{cases} \qquad (8)$$

Based on the post-Newtonian parameters (Will, 1981), in the general relativity the post-Newtonian parameters $\alpha_1 = \alpha_2 = \alpha_3 = 0$, $\beta = 1$, $\gamma = 1$, $\zeta_2 = 0$. and in the Brans-Dicke gravitational theuries $\alpha_1 = \alpha_2 = \alpha_3 = 0$, $\varsigma_2 = 0$ , β = 1, $\gamma = \dfrac{1+\omega}{2+\omega}$, ω is the dimensionless constant of the theory, ω = 5 ( Estabrook, 1969: Nordtvedt, 1970b )

3. The post-Newtonian perturbing equations and the perturbing variables

Substituting the perturbing accelerations R, S, W for the formulas (3), into the following Gaussian equations (Brouwer & Clemence, 1961)

$$\begin{cases} \dfrac{da}{dt} = \dfrac{2}{n(1-e^2)^{1/2}}(R\sin f + S\dfrac{p}{r}), \\ \dfrac{de}{dt} = \dfrac{(1-e^2)^{1/2}}{na}[R\sin f + S(\cos u + \cos f)], \\ \dfrac{dI}{dt} = \dfrac{1}{na(1-e^2)^{1/2}} W\dfrac{r}{a}\cos(\omega + f), \\ \sin I\dfrac{d\Omega}{dt} = \dfrac{1}{na(1-e^2)^{1/2}} W\dfrac{r}{a}\sin(\omega + f), \\ \dfrac{d\tilde{\omega}}{dt} = \dfrac{(1-e^2)^{1/2}}{nae}[-R\cos f + S(\dfrac{p}{r}+1)\sin f] + 2\sin^2\dfrac{1}{2}I\dfrac{d\Omega}{dt}, \\ \dfrac{d\varepsilon}{dt} = -\dfrac{2r}{na^2}R + \dfrac{e^2}{1+(1-e^2)^{1/2}}\dfrac{d\omega}{dt} + 2(1-e^2)^{1/2}\sin^2\dfrac{1}{2}I\dfrac{d\Omega}{dt}. \end{cases} \tag{9}$$

Here u is the eccentric anomaly. $\tilde{\omega}$ is the longitude of periastron and $\omega$ is the argument of periastron. $\varepsilon$ is the mean longitude at epoch.

We obtain the set of the post-Newtonian Perturbing equations

$$\begin{cases} \dfrac{da}{dt} = \dfrac{2}{n\sqrt{1-e^2}\,p}(\dfrac{1}{r^2})\Big\{[K_1 + \dfrac{3}{4}K_2 e^2 + (1+\dfrac{1}{4}e^2)K_3 + (1+\dfrac{1}{4}e^2)K_4]e\sin f \\ \qquad + \dfrac{1}{2}(K_1 + 2K_3 + 2K_4)e^2\sin 2f + \dfrac{1}{4}(K_3 - K_2 + K_4)e^2\sin 3f\Big\}, \end{cases}$$

$$\frac{de}{dt} = \frac{\sqrt{1-e^2}}{nap}(\frac{1}{r^2})\Big\{[K_1 + \frac{3}{4}e^2K_2 + (1+\frac{1}{4}e^2)K_3 + \frac{5}{4}e^2K_4]\sin f$$

$$+(\frac{1}{2}K_1 + K_3 + K_4)e\sin 2f + \frac{1}{4}(K_3 - K_2 + K_4)e^2\sin 3f\Big\}, \qquad \textbf{(10)}$$

$$\frac{d\tilde{\omega}}{dt} = \frac{\sqrt{1-e^2}}{naep}(\frac{1}{r^2})\Big\{(K_4 - \frac{1}{2}K_1 - K_3)e + [\frac{1}{4}e^2K_4 - (\frac{3}{4}e^2+1)K_3 - \frac{1}{4}e^2K_2 - K_1]\cos f$$

$$-(\frac{1}{2}K_1 + K_3 + K_4)e\cos 2f - \frac{1}{4}(K_4 + K_3 - K_2)e^2\cos 3f\Big\},$$

$$\frac{dI}{dt} = \frac{d\Omega}{dt} = 0,$$

$$\frac{d\varepsilon}{dt} = -\frac{2}{na^2}(\frac{1}{r^2})[(K_1 + K_2 + K_3) + K_2(e^2-1)\frac{r}{p} + (K_3 - K_2)e\cos f]$$

$$+(1-\sqrt{1-e^2})\frac{d\varpi}{dt}.$$

In the set of the equations (6) we transform independent variable time t into independent variable anomaly f by using $dt = r^2 df / na^2\sqrt{1-e^2}$ and $n^2a^3 = m$, and then, integrating the equations

$$\delta\sigma = \int_{\sigma_0}^{\sigma} d\sigma = \int_{f_0}^{f}[\frac{d\sigma}{dt}\frac{dt}{df}]df. \qquad (11)$$

In it $\sigma$ denotes arbitrary orbital elements from $a$, $e$, $\omega$, $i$, $\Omega$, and $\varepsilon_0$

Substituting $\frac{d\sigma}{dt}$ for the set of equations (6) and $\frac{dt}{df} = r^2 / na^2\sqrt{1-e^2}$ into the above definite integral expressions and integrating, one obtain the perturbation variables

$$\delta a = -\frac{2}{m(1-e^2)^2}\Big\{[K_1 + \frac{3}{4}e^2K_2 + (1+\frac{1}{4}e^2)K_3 + (1+\frac{1}{4}e^2)K_4]e(\cos f - \cos f_0)$$

$$+\frac{1}{4}(K_1+2K_3+2K_4)e^2(\cos 2f-\cos 2f_0)+\frac{1}{12}(K_3-K_2+K_4)e^3(\cos 3f-\cos 3f_0)\Big\},$$

$$\delta e=-\frac{1}{mp}\Big\{[K_1+\frac{3}{4}e^2K_2+(1+\frac{1}{4}e^2)K_3+\frac{5}{4}e^2K_4](\cos f-\cos f_0)$$

$$+\frac{1}{4}(K_1+2K_3+2K_4)e(\cos 2f-\cos 2f_0)+\frac{1}{12}(K_3-K_2+K_4)e^2(\cos 3f-\cos 3f_0)\Big\},$$

$$\delta I=\delta\Omega=0, \qquad \textbf{(12)}$$

$$\delta\tilde{\omega}=\frac{1}{mpe}\Big\{(K_4-\frac{1}{2}K_1-K_3)e(f-f_0)$$

$$+[\frac{1}{4}e^2K_4-(\frac{3}{4}e^2+1)K_3-\frac{1}{4}e^2K_2-K_1](\sin f-\sin f_0)$$

$$-\frac{1}{2}(\frac{1}{2}K_1+K_3+K_4)e(\sin 2f-\sin 2f_0)$$

$$-\frac{1}{12}(K_4+K_3-K_2)e^2(\sin 3f-\sin 3f_0)\Big\},$$

$$\delta\varepsilon=\frac{1}{mpe}\Big\{[(K_4-K_3-\frac{1}{2}K_1)-\sqrt{1-e^2}(\frac{3}{2}K_1+2K_2+K_3+K_4)]e(f-f_0]$$

$$+2(1-e^2)eK_2(u-u_0)+(1-\sqrt{1-e^2})[\frac{1}{4}e^2K_4-\frac{3}{4}(e^2+1)K_3-\frac{1}{4}e^2K_2-K_1]$$

$$-2e^2\sqrt{1-e^2}(K_3-K_2)](\sin f-\sin f_0)$$

$$-\frac{1}{2}(1-\sqrt{1-e^2})(\frac{1}{2}K_1+K_3+K_4)e(\sin 2f-\sin 2f_0)\ - \qquad \textbf{(13)}$$

$$-\frac{1}{12}(1-\sqrt{1-e^2})(K_4+K_3-K_2)e^2(\sin 3f-\sin 3f_0)\Big\},$$

$\delta l=n(t-t_0)+\delta\varepsilon.$ u is the eccentric anomaly

Here $l$ denotes the mean longitude of periastron

The perturbative solutions (12) and (13) of Keplerian elements include over ten kinds of gravitational theories as shown in Table 5.1 ( Will, 1981 ). Hence the formals (12)-(13) are important and worth-while.. In this paper we only select two kind theories of general relativity and Brans-Dicke

In the above last integral expression, we have used already the next integral expressions

$$\int r df = a\int \sqrt{1-e^2} du$$

$$\int \frac{e^2 \sin^2 f}{p} r df = \frac{(e^2-1)}{p}\int r df + \int df - e\int \cos f df.$$

Here u is the eccentric anomaly.

4. The secular variations of the orbital elements

It is seen from the results of the integration (8) that there exist the secular terms for $\delta\omega$ and $\delta\varepsilon_0$

$$(f - f_0) = n(t - t_0) + \ \ \textit{Periodic terms}$$

$$u - u_0 = n(t - t_0) + \ \ \textit{Periodic terms}$$

Here u denotes the eccentric anomaly and $u_0$ is the value of u as t=0

All other terms are the periodic terms for $\delta a$, $\delta e$, $\delta\tilde{\omega}$ and $\delta\varepsilon$. The coefficients of the periodic terms are the amplitudes of the periodic terms.

It is interesting for studying the secular terms. Hence we take the secular terms from the expressions (8) or integrating the definite integration (7) and taking the lower limit $f_0 = 0$ and upper limit $f = 2\pi$, the results of integration are that the periodic terms disappear and the secular terms appear per cycle by letting $m = GM / c^2$, we get

$$\Delta a = \Delta e = \Delta I = \Delta\Omega = 0,$$

$$\Delta\tilde{\omega} = -\frac{2\pi}{mp}(K_4 - K_3 - \frac{1}{2}K_1)\ {}^{rad}\!/_{cycle},$$

$$\Delta\varepsilon = -\frac{2\pi}{mp}[K_4 - K_3 - \frac{1}{2}K_1 + 2(1-e^2)K_2 - \tag{14}$$

$$- \sqrt{1-e^2}\,(\frac{3}{2}K_1 + 2K_2 + K_3 + K_4)]\, {}^{rad}\!/_{cycle}.$$

$$\Delta l = 2\pi + \Delta\varepsilon.$$

$$\Delta\tilde{\omega} = \Delta\omega + \Delta\Omega = \Delta\omega$$

The time variation of periastron passage, $\tau$, can be derived from the following relation

$$M_0 = \varepsilon - \omega - \Omega, \qquad M_0 = -n\tau$$

$$-n\Delta\tau - \tau\Delta n = \Delta\varepsilon - \Delta\omega - \Delta\Omega \tag{15}$$

$$\Delta n = -\frac{3}{2}(\frac{n}{a})\Delta a, = 0 \qquad \Delta\Omega = 0, \qquad n = \frac{2\pi}{P}$$

$$\Delta\tau = -\frac{1}{n}(\Delta\ \varepsilon - \Delta\omega) = -\frac{P}{2\pi}(\Delta\varepsilon - \Delta\omega)$$

The secular rates per year are

$$\begin{cases} \dot{a} = \dot{e} = \dot{I} = \dot{\Omega} = 0, \\ \dot{\tilde{\omega}} = \Delta\tilde{\omega}/P(rad/yr), \\ \dot{\varepsilon} \ = \Delta\varepsilon/P(rad/yr), \\ \dot{l} = 2\pi/P + \Delta\varepsilon_0/P(rad/yr), \\ \dot{\tau} = \Delta\tau/P(s/yr). \end{cases} \tag{16}$$

Here P denotes the orbital period, in yr.

Substituting $K_1$, $K_2$, $K_3$ and $K_4$ for the expressions ( 9 ) into the expressions (14) and by replacing $G$ and $c^2$ , then, we obtain the formulas for the secular variables and the variable rate in the general relativity

$$\begin{cases} \Delta\tilde{\omega}_{GR} = \dfrac{6\pi Gm}{c^2(1-e^2)}(rad/cycle), \\ \Delta\varepsilon_{GR} = \dfrac{6\pi Gm}{c^2a(1-e^2)}[1 + 2(1-e^2) - 5\sqrt{1-e^2}] + \dfrac{2\pi Gm}{c^2a(1-e^2)}[9\sqrt{1-e^2} - 7(1-e^2)](rad/cycle) \end{cases}$$

$$\Delta\tau_{GR} = -\frac{1}{n}(\Delta\varepsilon_{GR} - \Delta\varpi_{GR}) = -\frac{P}{2\pi}(\Delta\varepsilon_{0Gr} - \varpi_{Grr})(s/cycle)$$

(17)

$$\begin{cases} \dot{\tilde{\varpi}}_{GR} = \Delta\tilde{\varpi}_{GR} \; / P(rad/yr), \\ \dot{\varepsilon}_{GR} = \Delta\varepsilon_{GR} \; / P(rad/yr), \\ \dot{\tau}_{GR} = \Delta\tau_{GR} / P(s/yr). \end{cases}$$

(18)

In the Brans-Dicke theory

$$\begin{cases} \Delta\dot{\tilde{\varpi}}_{B-D} = \frac{38}{7}\frac{\pi Gm}{c^2(1-e^2)}(rad/cycle), \\ \Delta\varepsilon_{B-D} = \frac{38}{7}\frac{\pi Gm}{c^2 a(1-e^2)}[1+\frac{40}{19}(1-e^2)-\frac{99}{19}\sqrt{1-e^2}]+\frac{4\pi Gm}{c^2 a(1-e^2)}[6\sqrt{1-e^2}-5(1-e^2)](\frac{rad}{cycle \\ \Delta\tau_{B-D} = -\frac{P}{2\pi}(\Delta\varepsilon - \Delta\varpi)(s/cycle) \end{cases}$$

(19)

$$\begin{aligned} \dot{\tilde{\varpi}}_{B-D} &= \frac{\Delta\tilde{\varpi}_{B-D}}{P}(Rad/yr) \\ \dot{\varepsilon}_{B-D} &= \frac{\Delta\varepsilon_{B-D}}{P}(rad/yr) \\ \dot{\tau}_{B-D} &= \frac{\Delta\tau_{B-D}}{P}(s/yr) \end{aligned} \qquad (20)$$

5. Numerical calculation for six eexotrasolar planets

In this paper we choose six exoplanets: HD16871b, HD68988b, HD217107b, HD88133b, XO-3b and GJ-436b as an example For the former exoplanets, their P(d), $M^*(M_\odot)$ are retrieved from Bodenheimer et al (2003) and a(Au), e and m $_b$ (m $_J$) are cited from www.mpia.de/homes/Lyra/planet_naming.html.; For latter three Exoplanets, their P(d) a(Au), $M^*(M_\odot)$ and e are retrieved from Jordan & Bakos (2008) and m $_b$(m $_J$) is cited from http://www.exoplanet.eu/index.php. These data are listed in Table 5 of the Appendix.

Substituting those data into formulas (17) – (20), we obtain the numerical results for the secular variation of the orbital element of six exoplanets in Table1 and Table 2.

Table 1 The secular variations per revolution for the orbital elements of six extrasolar planets

| exoplanets | $\Delta\tilde{\varpi}_{GR}$ ("/Rev) | $\Delta\varepsilon_{GR}$ ("/Rev) | $\Delta\tilde{\omega}_{B-D}$ ("/Rev) | $\Delta\varepsilon_{B-D}$ ("/Re) | $\Delta\varepsilon_{GR}$ (s/yr) | $\Delta\tau_{B-D}$ (s/yr) |
|---|---|---|---|---|---|---|
| HD68988b | 0".61 | -0".78 | 0."55 | -0"73- | -0.58 | 0.54 |
| HD16874b | 0.62 | -0.81 | 0.56 | --0.76 | -0.61- | 0.56 |
| HD217107b | 0.52 | -0.68 | 0.47 | -0.64 | 0.57 | 0.53 |
| HD88133b | 0.99 | -1.32 | -0.90 | --1.20 | -0.52- | 0.48 |
| XO-3b | 1.24 | -1.51 | 1.10 | -1.36 | -0.57- | 0.52 |
| GJ-436b | 0.60 | -0.74 | 0..52 | -0.69 | -0.23 | 0.21 |

Note: " denotes are second and: Rev denotes Revolution (Cycle)

Table2. The secular rates per year for the orbital elements of six Extrasolar Planets

| exoplanets | $\dot{\tilde{\varpi}}_{GR}$ ("/yr) | $\dot{\varepsilon}_{GR}$ ("/yr) | $\dot{\omega}_{B-D}$ ("/yr) | $\dot{\varepsilon}_{B-D}$ ("/yr) | $\dot{\varepsilon}_{GR}$ (s/yr) | $\dot{\tau}_{B-D}$ (s/yr) |
|---|---|---|---|---|---|---|
| HD68988b | 35".61 | -45.39 | 31.18 | -42.66 | 33.84 | 31.16 |
| HD16874b | 35.19 | --46.60 | 31.83 | --43.18 | 34.81 | 32.14 |
| HD217107b | 26.86 | -35.22 | 24.29 | -33.01 | -29.21 | 27.02 |
| HD88133b | 106.64 | -153.94 | 96.76 | -128.73 | 56.24 | 51.13 |
| XO-3b | 142.46 | -176.26 | 125.86 | -155.30 | -66.22 | -59.89 |
| GJ-346b | 83.19 | -103.19 | 71.97 | -95.04 | -31.89 | -29.46 |

6.Discussion (1)

Comparison with the results of other authosr

The results of the numerical values of advance of periastron of HD88133b, XO-3b and GJ-436b in this paper as compared with that of three exoplanets in the other author's work ( Jord$a'$n & Bakos, 2008) are listed in Table3.

Table3. Comparison with other authors for three exoplanets

| exoplanets | This study | | ordan & Bakos (2008) |
|---|---|---|---|
| | $\dot{\tilde{\varpi}}$ (″/cy) | $\dot{\tilde{\varpi}}$ (deg/cy) | $\dot{\tilde{\omega}}$ (deg/cy) |
| HD88133b | 10664″ | 2°.961 | 2°.958 |
| XO-3b | 14246 | 3°.959 | 3°.886 |
| GJ-436b | 8319 | 2°.311 | 2°.234 |

We can seen from the above Table 3 that both results are nearly approximate in the relativistic effect, but there are some different . The calculated values of this paper are some larger than that of Jordán & Bakos (2008) This difference results in that this paper calculates $\dot{\omega}_{GR}$ by using the mass of two-body ( parent star and exoplanet ) and Jordan & Bakos only consider the mass of the parent star and neglect the mass of exoplanet.

(2) Comparison with the planets in solar system

Substituting the data of Mercury and Jupiter into the formulas (11) for $\Delta\tilde{\omega}_E$, we obtain the results for the comparison of the perihelion of Mercury and Jupiter per century with that of two exoplanets per century listed in Table 3

Table 4 The results of comparison with that of the Mercury and Jupiter

| exoplanets | $(\Delta\tilde{\omega})_{\max}$ | $(\Delta\tilde{\varpi})_{\min}$ | Planets in solar system | $(\Delta\tilde{\varpi})_{\max}$ | $(\Delta\tilde{\varpi})_{small}$ |
|---|---|---|---|---|---|

| | ( “/cy ) | ( ″ /cy ) | | ( ″ /cy ) | ( ″ /cyt ) |
|---|---|---|---|---|---|
| XO-3b | 14246″ | | Mercury | 42″ .91 | |
| HD217107b | | 2686″ | Jupiter | | 0″ .06 |

We can see from Table 4 that the values of advance of the periastron of the exoplanets are largest than that of the planets in the solar system. Therefore, it is important and meaningful for studying the motion of the Exoplanets.

(3) On the possibility of observing these effects.

Let us discuss the possibility of observing these effects. In the solar system the advance of perihelion of Mercury is 42 “.91 per century. We may see from Table 3 that in the extrasolar planetary system the maximal value of advance of XO-3b is 14246″ per century which correspond to 332 time ( fold ) value of advance of perihelion of the Mercury. At present, some authors applied the method of TTV (transit timing variation ) or the method of TDV( transitduration variation ). i.e, the secular precession can be detected through the long-term change in $P_{obs}$ or in $T_D$ ( TDV) to the observation of the extrasolar planetary system ( Agol et al (2005), Rabus et al (2009), Gibson et al (2009) , Iorio ( 2011b ).. Therefore, the possibility that the non-Newtonian advances of the periastra of the extrasolar planets considered can be observed is certainly interesting and deserves further studies

(4) The author emphasezes that when we consider slouly orbiting planers, we could look like secular term over relatively short observational time interval, i. e, the relatively short arcs or the short term effect are available and important. Hence the author takes the time interval per year in the Table 1 and Table 2.

( 5 ) Prospect for further investigation ( the new try )

At present, the fifth force, Yukawa-like interaction has been investigated in our solar system ( Iorio, 2007 b, Haranes & Ragol 2011, Tsang 2012 ). It may be predicted that extrasolar planets may well be used also for constraining putative fifth force, Yukawa-like interaction in the futher investigation.

7. Conclusions

In this paper we worked out parameterized post-Newtonian effect on the orbits of celestial objects. The semi-major axis and eccentricity exhibit periodic variation, but no secular variation. The longitude of periastron and mean longitude at epoch exhibit secular and periodic variation. The inclination and the longitude of ascending node are unaffected. Such effects on the orbits of the extrasolar planets may be observed possibly because their effects of advance of periastron are large as in the calculation of this paper. The results of this paper based on the parameterized post-Newtonian gravitational metric by the work of C M Will , amplified and extended his work.

Appendix

Table 5 Orbital and physical parameters of six extrasolar planetary systems used in the text

| exoplanets | P(d) | *a* (Au) | $M^*(m_\odot)$ | $m_b(m_J)$ | e | Ref |
|---|---|---|---|---|---|---|
| HD68988b | 6.276 | 0.071 | 1.11 | 1.90 | 0.140 | Bodenheimer et al (2003) |
| HD16874b | 6.403 | 0.065 | 1.09 | 0.23 | 0.081 | Bodenheimer] et al (2003) |
| HD217107b | 7.125 | 0.073 | 0.98 | 1.33 | 0.132 | [Bodenheimer et al (2003) |
| HD88133b | 3.416 | 0.047 | 1.20 | 0.22 | 0.133 | Jordan Bakos (2008) |
| XO-3b | 3.192 | 0.048 | 1.41 | 11.79 | 0.260 | Jordan Bakos (2008) |
| GJ-436b | 2.644 | 0.028 | 0.41 | 0.0737 | 0.159 | Jordan Bakos (2008) |

$m_b$ denotes exoplanet mass which is cited from http://www.mpia.de/homes/Lyra/planet_naming.html for the former three references, and the latter three references is cited from http://www.exoplanet.eu/index.php for $m_b$.

## .References